\documentclass[aps,prl,preprint,superscriptaddress]{revtex4-2}
\usepackage{amssymb}
\usepackage{amsmath}
\usepackage{graphicx}
\usepackage{graphicx}
\usepackage{combelow}
\usepackage[utf8]{inputenc}
\usepackage{color}
\usepackage{textcomp} 
\usepackage{bm}       
\usepackage{float}

\usepackage{epstopdf}
\AppendGraphicsExtensions{.tif}

\begin{document}
	
	\title{Tuning the coexistence regime of incomplete and tubular skyrmions in ferro/ferri/ferromagnetic trilayers}
	
	\author{O\u{g}uz Y{\i}ld{\i}r{\i}m}
	\email{oguz.yildirim@empa.ch: yildirim$_$o@outlook.com}
	\affiliation{Empa, Swiss Federal Laboratories for Materials Science and Technology, CH-8600 Dübendorf, Switzerland}
	\author{Riccardo Tomasello}
	\affiliation{Department of Electrical and Information Engineering, Politecnico di Bari, 70125 Bari, Italy}
	\author{Yaoxuan Feng}
	\affiliation{Empa, Swiss Federal Laboratories for Materials Science and Technology, CH-8600 Dübendorf, Switzerland}
	\author{Giovanni Carlotti}
	\affiliation{Dipartimento di Fisica e Geologia, Università di Perugia, 06123 Perugia, Italy}
	\author{Silvia Tacchi}
	\affiliation{Istituto Officina dei Materiali - IOM, 06123 Perugia, Italy}
	\author{Pegah Mirzadeh Vaghefi}
	\affiliation{Empa, Swiss Federal Laboratories for Materials Science and Technology, CH-8600 Dübendorf, Switzerland}
	\author{Anna Giordano}
	\affiliation{Department of Mathematical and Computer Sciences, Physical Sciences and Earth Sciences, University of Messina, I-98166 Messina, Italy}
	\author{Tanmay Dutta}
	\affiliation{Empa, Swiss Federal Laboratories for Materials Science and Technology, CH-8600 Dübendorf, Switzerland}
	\author{Giovanni Finocchio}
	\affiliation{Department of Mathematical and Computer Sciences, Physical Sciences and Earth Sciences, University of Messina, I-98166 Messina, Italy}
	\author{Hans J. Hug}
	\affiliation{Empa, Swiss Federal Laboratories for Materials Science and Technology, CH-8600 Dübendorf, Switzerland}
	\affiliation{Department of Physics, University of Basel, CH-4056 Basel, Switzerland}
	\author{Andrada-Oana Mandru}
	\email{andrada-oana.mandru@empa.ch}
	\affiliation{Empa, Swiss Federal Laboratories for Materials Science and Technology, CH-8600 Dübendorf, Switzerland}


\begin{abstract}
The development of skyrmionic devices requires a suitable tuning of material parameters in order to stabilize skyrmions and control their density. It has been demonstrated recently that different skyrmion types can be simultaneously stabilized at room temperature in heterostructures involving ferromagnets, ferrimagnets and heavy metals, offering a new platform of coding binary information in the type of skyrmion instead of the presence/absence of skyrmions. Here, we tune the energy landscape of the two skyrmion types in such heterostructures by engineering the geometrical and material parameters of the individual layers. We find that a fine adjustment of the ferromagnetic layer thickness and thus its magnetic anisotropy, allows the trilayer system to support either one of the skyrmion types or the coexistence of both and with varying densities.
\end{abstract}

\clearpage

\maketitle 

\section{Keywords}

Magnetic skyrmions, magnetic multilayers, skyrmion lattice, skyrmion type, magnetic force microscopy

\section{INTRODUCTION}
Magnetic skyrmions, local spin textures possessing topological protection and particle-like nature \cite{Bogdanov1989, Bogdanov1994}, have been extensively studied in a broad range of materials including bulk chiral magnets \cite{Muhlbauer2009}, ferromagnetic (FM) thin films grown on single crystals \cite{Heinze2011, Romming2015}, polycrystalline multilayer systems \cite{Chen2015, Moreau-Luchaire2016, Soumyanarayanan2017}, ferrimagnets and, more recently, synthetic antiferromagnets \cite{Dohi2019, Legrand2020}. Even though the field of skyrmions is relatively young, a tremendous amount of progress has been made in terms of both skyrmion stabilization and their dynamics. For the specific case of thin film multilayered materials, in less than one decade, they evolved from hosting atomic-scale skyrmions, but stable in very high fields at low temperature \cite{Heinze2011, Romming2015} to nanometer-sized skyrmions, stabilized by low magnetic fields at room temperature by strategically utilizing interfacial effects such as the Dzyaloshinskii-Moriya interaction (DMI) between a FM layer and an adjacent heavy metal \cite{Moreau-Luchaire2016, Soumyanarayanan2017}.

Due to their small size and possibilities of easily transporting them, room temperature magnetic skyrmions may serve as information carriers in very compact and also energetically-efficient storage such as racetrack memory \cite{Parkin2008, Fert2013, Tomasello2014S, Yu2017}. There are however two fundamental limitations to using skyrmions as magnetic bits in such a device: 1) Not being able to restrict their motion along straight paths due to the skyrmion Hall effect, leading to skyrmions being lost at the device edge \cite{Jiang2017, Litzius2017}; one possible solution is the use of antiferromagnetically-coupled skyrmions \cite{Dohi2019, Legrand2020}; and 2) Not having stable inter-skyrmion distances, leading to fluctuating distances among bits; a potential solution was revealed with the experimental observation of coexisting skyrmions and chiral bobbers in B20-type crystalline materials at low temperatures \cite{Zheng2018}, alluding to the possibility that a chain of binary data bits could be encoded by two different soliton states. Ideally however, the two states would appear at room temperature in multilayer systems that allow an immediate integration in current device technology. 

In our previous study \cite{Mandru2020}, we developed a FM/Ferrimagnetic/FM (FM/FI/FM) trilayer system that can host two distinct skyrmion types at room temperature and can serve as a solution to the second limitation. The FM Ir/Fe/Co/Pt (multi)layers generate a high density of small N{\'e}el skyrmions by means of additive interfacial DMI arising from the Ir/Fe and Co/Pt interfaces \cite{Soumyanarayanan2017}. The role of the FI layer is to provide perpendicular anisotropy and a small DMI, but cannot otherwise host small skyrmions or form small perpendicular domains. Due to these special properties of the FI, the skyrmions that are generated by the bottom and top FM layers can penetrate through the FI, leading to the stabilization of tubular skyrmions (running through the entire trilayer). The ones that do not penetrate through remain as incomplete skyrmions (existing in the top and bottom FM layers only). Both types of skyrmions can be stabilized within the same sample, as revealed from magnetic force microscopy (MFM) data and micromagnetic simulations.  
	
Having established such a trilayer as a platform for hosting two skyrmion types, its implementation into memory devices and even beyond (e.g., logic devices \cite{Zhang2015}, unconventional computing architectures \cite{Prychynenko2018, Pinna2018} and transistor applications \cite{Zhang2015_2}) requires tunable systems and good control over the bit type and density. To this end, we explore the magnetic parameter space of the individual layers within the trilayer structure, revealing that the coexistence range of the two skyrmions can be tuned by altering the magnetic properties of the FM layers. More specifically, we investigate the trilayer system when changing the thickness of the FM layers by using a combination of MFM, vibrating sample magnetometry (VSM) and Brillouin light scattering (BLS) experiments together with micromagnetic simulations. We show how the competition between different magnetic energies present in this system allows for the stabilization of either incomplete or tubular skyrmions, or a combination of both with varying ratios, providing a useful avenue for tuning the skyrmion type and density within the same material. Having a good understanding of the interplay between different magnetic properties can lead to better-tailored systems and possibly bring skyrmions a step closer to applications. 

\section{RESULTS AND DISCUSSION}
\begin{figure}[t]
	\includegraphics[width=0.7\textwidth]{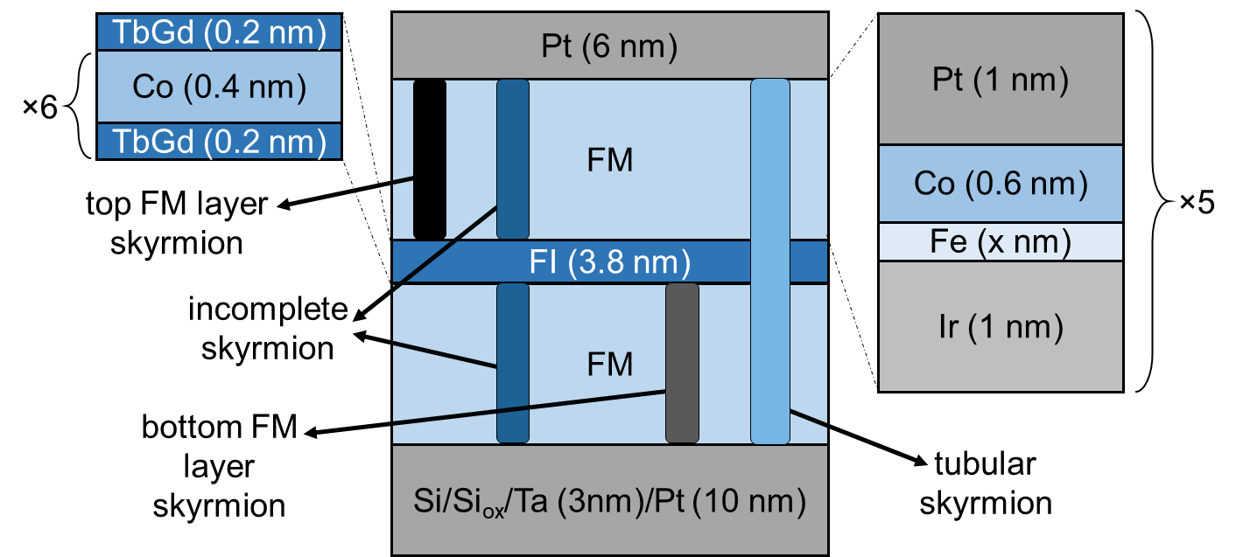}
	\caption{[Ir/Fe($x$)/Co/Pt]$_{\times 5}$/[(TbGd)/Co]$_{\times 6}$/(TbGd)]/[Ir/Fe($x$)/Co/Pt]$_{\times 5}$ FM/FI/FM trilayer sample schematics with $x$ being the Fe-sublayer thickness that is modified among different samples. The various skyrmion types that could appear in such samples are also indicated: top and bottom FM layer skyrmions along with incomplete (in both FM layers) and tubular (in all FM and FI layers) skyrmions.}
	\label{fig:Fig1} 
\end{figure}

The FM/FI/FM trilayer schematics is shown in Figure \ref{fig:Fig1}. The samples consist of a FI [(TbGd)(0.2)/Co(0.4)]$_{\times 6}$/(TbGd)(0.2)] layer sandwiched between two skyrmion-generating FM [Ir(1)/Fe($x$)/Co(0.6)/Pt(1)]$_{\times 5}$ layers; all nominal thicknesses in parentheses are in nm and $x$ is the Fe-sublayer thickness, the only parameter that is varied for the different samples. The Fe was varied since it has been previously shown that thickness changes in this particular sublayer leads to a more dramatic change in the skyrmion density as opposed to changes in the Co sublayer \cite{Soumyanarayanan2017}. The thickness and magnetic properties of the FI layer are kept the same for all samples (see Figure S1 in the supplementary information).

In order to establish a base for discussion, we next summarize our earlier results \cite{Mandru2020} on a sample similar to that shown in Figure \ref{fig:Fig1} having an Fe thickness $x = 0.29\,$nm and with comparable magnetic properties for the FI layer (see section 1 in the supplementary information). Zero-field MFM investigations on this sample revealed, as the lowest energy configuration, a maze domain pattern with two different types of domains having lower and higher frequency shift $|\Delta f|$ contrast. In an increasing applied magnetic field, these two domain types give rise to two distict skyrmions, also having lower and higher $|\Delta f|$ contrast and corresponding to smaller- and larger-diameter skyrmions, respectively. The locations of the two types of skyrmions throughout the trilayer structure were determined by preparing a subset of layers from the original sample and by performing MFM investigations using the same tip and the same tip-sample distance for all samples. One of the most relevant samples was a trilayer with non-magnetic Ta instead of the FI, i.e. FM/Ta(3.8)/FM, revealing that its MFM $|\Delta f|$ contrast matched best that of the low contrast skyrmion in the original FM/FI/FM structure, thus indicating that such an incomplete skyrmion exists in both the top and bottom FM layers, but not in the FI. Another relevant sample was the FM [Ir(1)/Fe(0.29)/Co(0.6)/Pt(1)]$_{\times 5}$ since it revealed, based on its much lower $|\Delta f|$ contrast, that the incomplete skyrmion contrast cannot correspond to skyrmions in the top FM layer of the FM/FI/FM structure. Regarding the high contrast skyrmions, no other subset of samples showed such a large contrast and therefore this type of skyrmion was attributed to a tubular skyrmion that was present in all three layers of the structure. The incomplete and tubular skyrmions are shown schematically in Figure \ref{fig:Fig1}. Using micromagnetic simulations, we found that such a trilayer structure can indeed host the two types of skyrmions and that one of the key ingredients required to stabilize a tubular skyrmion is to have sufficient DMI in the FI layer, namely $D_\text{FI}$ = 0.8\,mJ/m$^{2}$. Note that in addition to incomplete and tubular skyrmions, such samples could in principle also host only top or bottom FM layers skyrmions (also shown schematically in Figure \ref{fig:Fig1}). We have found top FM layer skyrmions in our earlier study that are very few in number (about 8\,\% of the total number of skyrmions in a given area, see the supplementary information of ref.\,\cite{Mandru2020}). However, we could not detect any bottom FM layer skyrmions: although the top FM layer skyrmions could easily be detected by MFM at a distance of 21\,nm (6\,nm Pt capping layer\,+\,15\,nm tip-sample distance) from the top-most FM layer, the bottom ones would not be so easily discernable due to the fact that the MFM signal decays exponentially with distance and they would have to be detected at 39.3\,nm (21\,nm\,+\,14.5\,nm\,+\,3.8\,nm), making the MFM experiments quite challenging. Nonetheless, if they exist, it is expected that their number is also very small compared to the total number of incomplete and tubular skyrmions observed in such trilayers.

\begin{figure}[t]
	\includegraphics[width=1\textwidth]{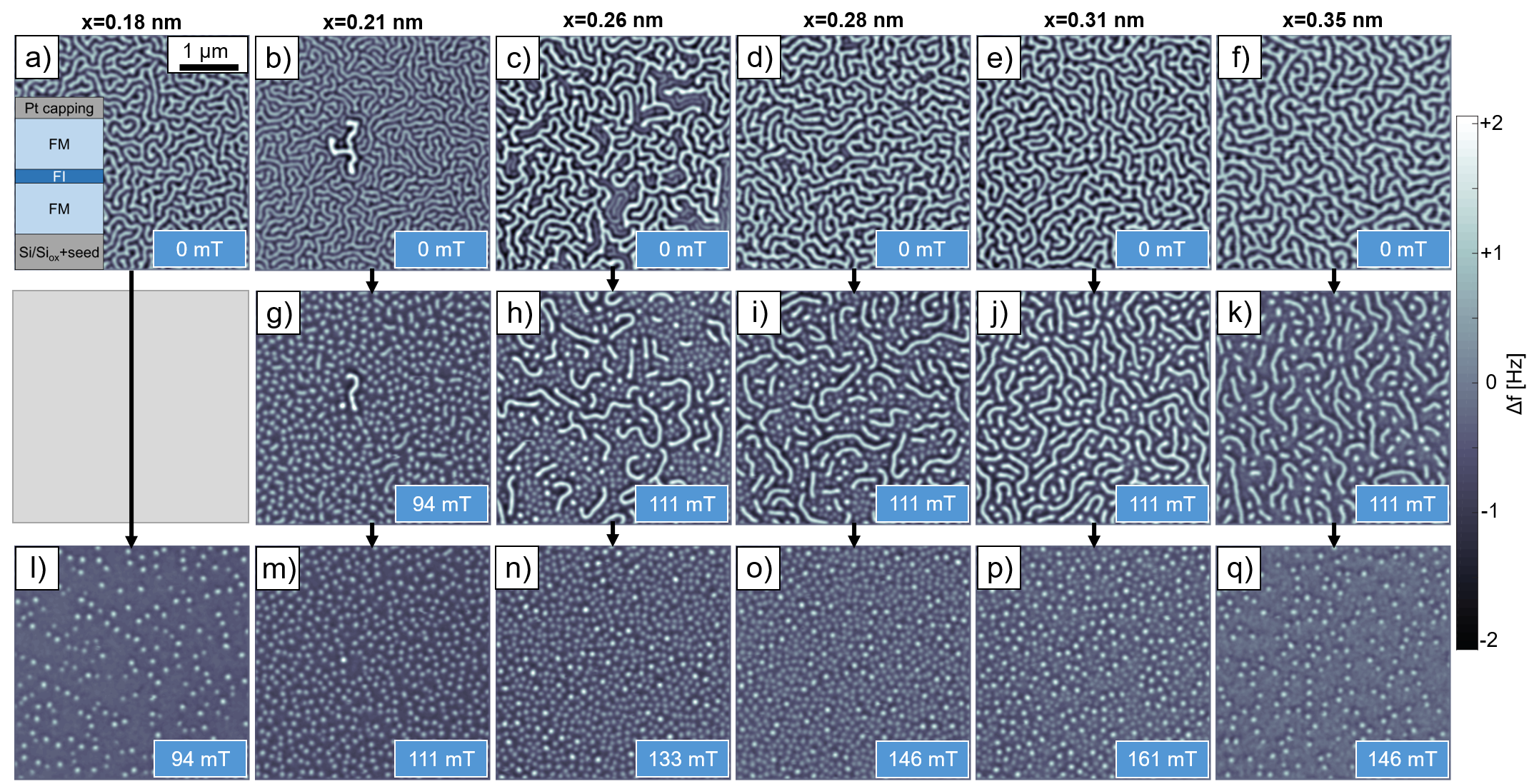}
	\caption{(a)-(f) 4\,{\textmu}m\,$\times$\,4\,{\textmu}m zero-field MFM images of the trilayer samples for five selected Fe-sublayer thicknesses, taken at remanence after out-of-plane saturation in 450\,mT magnetic field (above the saturation field of all samples, see Figure S2 in the supplementary information). (g)-(k) Intermediate-field MFM images taken on the same area as the images shown in (a)-(f); note that there is no intermediate-field MFM image for the $x = 0.18\,$ sample. (h)-(l) Subsequent MFM images recorded at the fields where skyrmions-only exist (no maze domains left). The inset from (a) shows a simplified schematics of the trilayer structure. All images were taken under the same conditions, with the same tip and at the same tip-sample distance. The $|\Delta f|$ contrast scale is the same for all images.}
	\label{fig:Fig2} 
\end{figure}

Figures \ref{fig:Fig2}(a)-(q) illustrate MFM images of the trilayer samples for six Fe-sublayer thicknesses in zero (top row), intermediate (middle row) and at the first fields where solely skyrmions are nucleated (bottom row). A common feature for all samples is that they show a maze domain pattern in zero field. Some of the samples however indicate the presence of two different types of domains having higher and lower $|\Delta f|$ contrast, as visible in Figs. \ref{fig:Fig2}(g)-(j) taken at intermediate magnetic fields, where apart from extended domains, some skyrmions appear. As the field is increased further, the domains shrink to the point where only skyrmions exist [Figs. \ref{fig:Fig2}(l)-(q)]. It is clear from Figs. \ref{fig:Fig2}(m)-(p) that the two different $|\Delta f|$ contrast domains from Figs. \ref{fig:Fig2}(g)-(j) give rise to two distinct skyrmion types. Figures \ref{fig:Fig2}(l) and (q) on the other hand indicate the nucleation of mainly single-contrast skyrmions.
\begin{figure}[t]
	\includegraphics[width=0.6\textwidth]{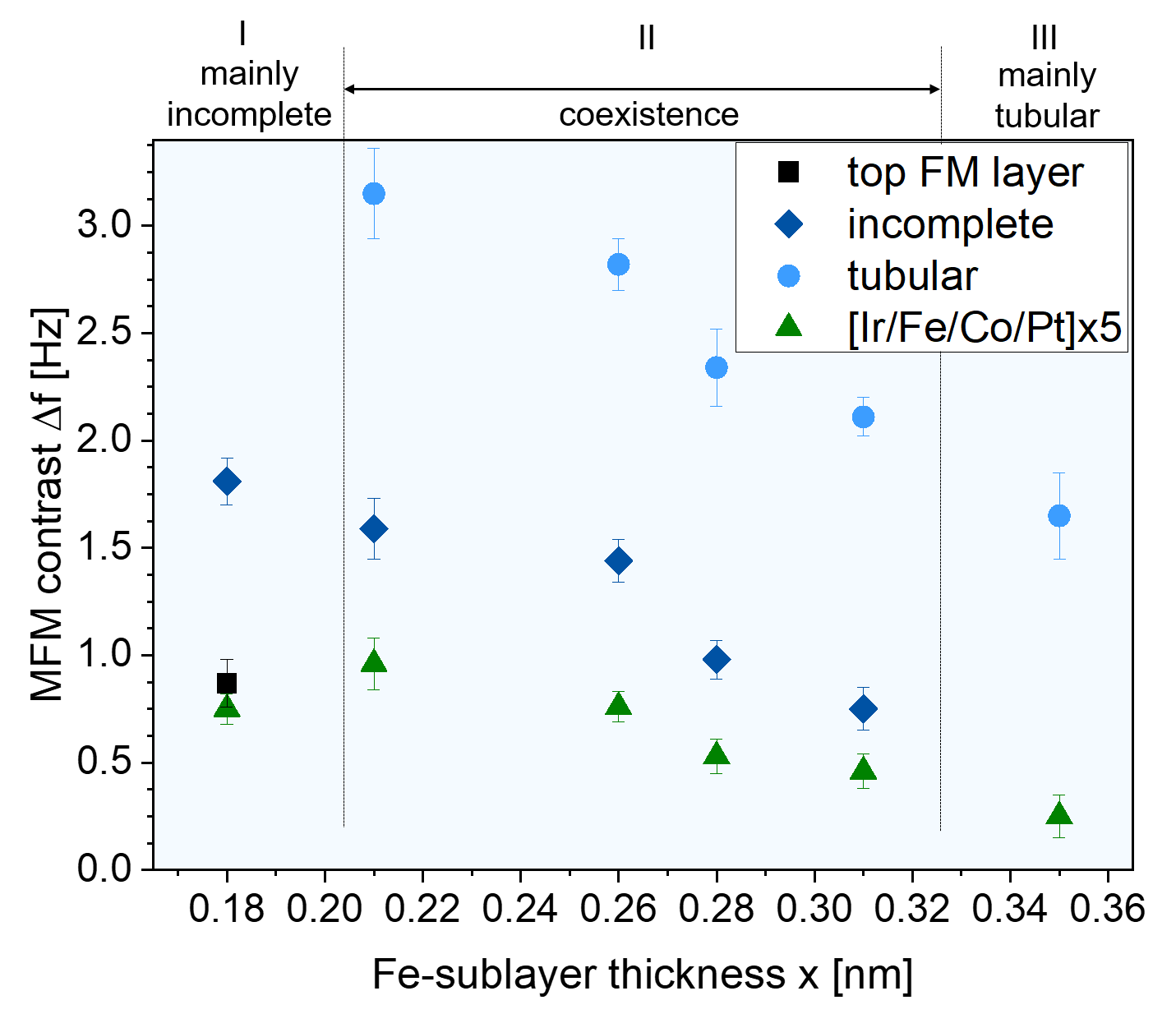}
	\caption{MFM $|\Delta f|$ contrast comparison between the three types of skyrmions observed in the trilayer samples and individual [Ir(1)/Fe($x$)/Co(0.6)/Pt(1)]$_{\times 5}$ samples as a function of Fe-sublayer thicknesses $x$. Regions I, II, and III correspond to experimental observations of the mainly incomplete-skyrmion sample, coexisting incomplete- and tubular-skyrmion samples, and mainly tubular-skyrmion sample. The contrast points for the tubular skyrmions in the $x = 0.18\,$nm sample and the incomplete skyrmions in the $x = 0.35\,$nm sample are not shown due to their small numbers compared to the other samples.}
	\label{fig:Fig3} 
\end{figure}

\begin{figure}[t]
	\includegraphics[width=\textwidth]{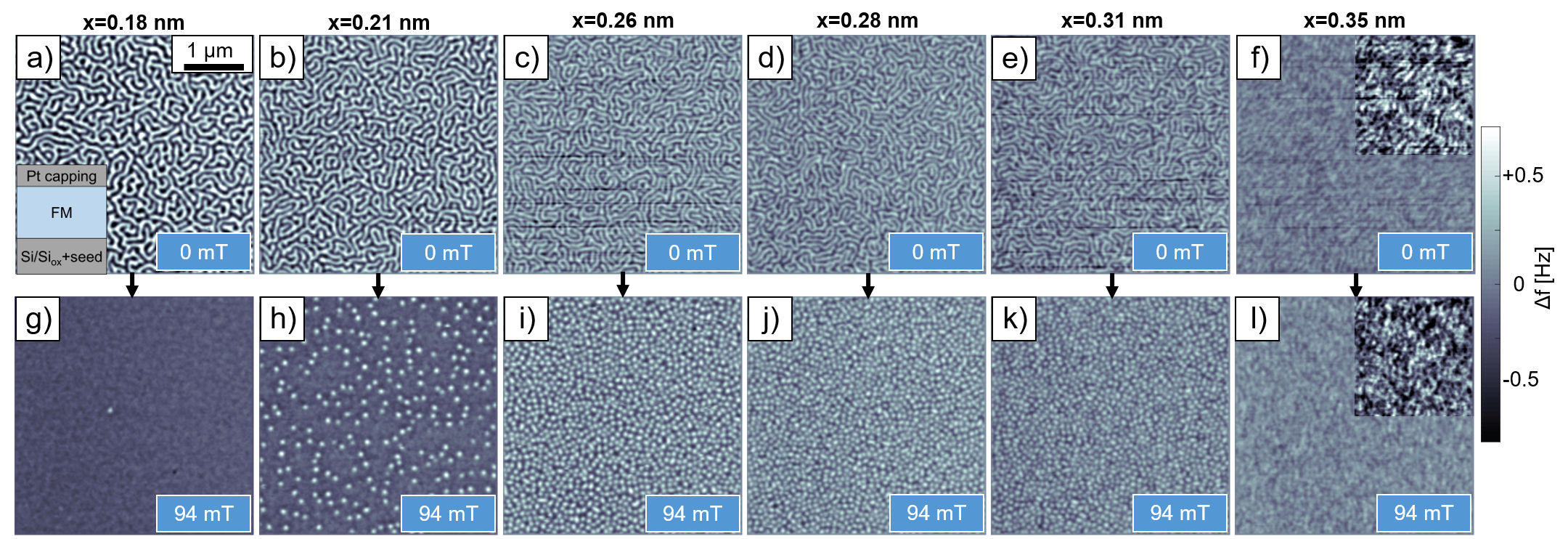}
	\caption{(a)-(f) 4\,{\textmu}m$\,\times$\,4\,{\textmu}m zero-field MFM images performed on individually-grown FM [Ir(1)/Fe($x$)/Co(0.6)/Pt(1)]$_{\times 5}$ layers for the same Fe thicknesses $x$ as in Figure \ref{fig:Fig2}; all data are taken at remanence after out-of-plane saturation in 450\,mT magnetic field. (g)-(l) Subsequent MFM images recorded at the fields where mainly skyrmions exist and no maze domains are left. The inset from (a) shows a simplified schematics of the FM layers. The $|\Delta f|$ contrast scale is the same for all 4\,{\textmu}m$\,\times$\,4\,{\textmu}m} images. The insets from f) and l) show enhanced-contrast (with $|\Delta f|$ scale 0.4\,Hz for both) data corresponding to 2\,{\textmu}m$\,\times$\,2\,{\textmu}m parts of the 4\,{\textmu}m$\,\times$\,4\,{\textmu}m images. All MFM data are taken with the same tip, at the same tip-sample distance and under the same conditions as the images from Figure \ref{fig:Fig2}.
	\label{fig:Fig4} 
\end{figure}

By comparing the MFM $|\Delta f|$ contrasts between different samples and also with a selected subset of layers grown for comparison, we can establish what types of skyrmions are present in our films following a similar line of experiments and argumets as for the initial study from ref.\,\cite{Mandru2020} described above. The MFM $|\Delta f|$ contrast comparison plot is presented in Figure \ref{fig:Fig3} and the analyses were performed on the data shown in Figs. \ref{fig:Fig2}(l)-(q) for the trilayers and in Figs. \ref{fig:Fig4}(g)-(l) for the corresponding FM [Ir(1)/Fe($x$)/Co(0.6)/Pt(1)]$_{\times 5}$ skyrmion-generating layers. Note that such a comparison is only possible under the same MFM imaging conditions, at the same tip-sample distance \cite{Zhao2018}, using the same tip and for the same capping-layer thickness among these samples (see Methods for further details). The highest contrast points (light blue symbols) correspond to tubular skyrmions and the lower contrast points (dark blue symbols) are attributed to incomplete skyrmions. The green symbols, corresponding to the [Ir(1)/Fe($x$)/Co(0.6)/Pt(1)]$_{\times 5}$ samples are also shown to either validate or discard the possibility of top FM layer skyrmions. Considering the $x = 0.28\,$nm sample [Figs. \ref{fig:Fig2}(d), (i) and (o)], analogous to the sample presented in our previous study for a very similar (0.29\,nm) Fe-sublayer thickness \cite{Mandru2020}, the two observed skyrmion contrasts correspond to an incomplete skyrmion (with lower $|\Delta f|$ and existing in the FM layers only) and a tubular skyrmion (with higher $|\Delta f|$ and running through the entire structure). Three other samples, i.e. with $x$ = 0.21, 0.26 and 0.31\,nm also show two types of domains/skyrmions that correspond to incomplete and tubular skyrmions. For these four samples that show the coexistence of the two skyrmion types, the density of both skyrmions increases with increasing Fe thickness. If the Fe-sublayer thickness is increased further to $x = 0.35\,$nm, we observe mainly one type of contrast. Considering the contrast evolution of both the tubular and incomplete skyrmions in the samples with smaller Fe thicknesses, we conclude that they are tubular skyrmions; these skyrmions have the lowest $|\Delta f|$ contrast among all samples that host tubular skyrmions. Note that this sample also has a very small number of incomplete skyrmions (not shown as a contrast data point in Figure \ref{fig:Fig3}). Finally, for the sample with the smallest Fe-thickness, i.e. $x = 0.18\,$nm, the MFM data and contrast analysis show incomplete skyrmions that have the lowest density out of all samples (except the one with $x = 0.35\,$nm). However, this sample shows another contrast (black symbol in Figure \ref{fig:Fig3}) that corresponds to a top FM skyrmion since it has about the same value as the FM [Ir(1)/Fe(0.18)/Co(0.6)/Pt(1)]$_{\times 5}$ sample (green symbol). At very isolated locations on the $x = 0.18\,$nm sample, we observe (not shown) a very small number of tubular skyrmions that have a higher contrast than those found in the $x = 0.21\,$nm sample, which is expected considering the contrast trends in Figure \ref{fig:Fig3}. We therefore identify three separate regimes for the stabilization of different skyrmion types as a function of the Fe-sublayer thickness: I) mainly incomplete skyrmions, II) coexistence of incomplete and tubular skyrmions, and III) mainly tubular skyrmions. 

The MFM $|\Delta f|$ contrast for all skyrmions (including those in the [Ir(1)/Fe($x$)/Co(0.6)/Pt(1)]$_{\times 5}$ samples - discussed in more detail below) decreases with increasing Fe-thickness. Due to the fact that both the magnetization and the total thickness change very little between these samples (see Figs. S2 and S3 in the supplementary information), the decay of the MFM contrast for all skyrmions with increasing Fe thickness can be attributed to a smaller skyrmion diameter.  This conclusion is reached by noting the relationship between the MFM contrast and skyrmion size and is explained as follows. The MFM contrast depends on the effective magnetic surface charge, the spatial wavelengths of the imaged features, the tip-sample distance, and on the tip transfer function (i.e. the decay of the measured frequency shift signal with decreasing spatial wavelength of the stray field at the location of the tip \cite{Feng2022}). Note that the stray field decays exponentially with the spatial wavelength and increasing tip-sample distance. Therefore, if the tip-sample distance is kept constant at $z\,\approx 12\,$nm for all measured samples \cite{Zhao2018} and the magnetization and total thickness of the samples are about the same, then the observed reduction in MFM skyrmion contrast can be attributed to a reduced skyrmion diameter. Note that the same argument also applies to the incomplete and tubular skyrmion diameters in a given sample: the total length of a tubular skyrmion is a factor of $\approx$\,1.1 larger than that of the incomplete skyrmion for the case of $x = 0.28\,$nm sample. Therefore, an increase of the contrast generated by the tubular versus the incomplete skyrmion in this sample by much more than this factor ($\approx$\,2.4) must thus arise from a larger diameter of the tubular skyrmion (that is also confirmed by micromagnetic simulations).

Since the properties of the FI layer are kept the same for all current samples, we conclude that the FM skyrmion-generator layers with varying Fe thickness are solely responsible for the three different regimes determined from MFM imaging and observed contrast evolution with Fe thickness. Therefore, along with MFM imaging, we have further investigated the individual [Ir(1)/Fe($x$)/Co(0.6)/Pt(1)]$_{\times 5}$ layers by extracting magnetic parameters for each sample from vibrating sample magnetometry (VSM) measurements and by measuring DMI constants using Brillouin light scattering (BLS) experiments (see Methods and Sections 3 and 4 in the supplementary information).

In terms of MFM results, varying the Fe-sublayer thickness has an impact on the magnetic properties of the FM [Ir(1)/Fe($x$)/Co(0.6)/Pt(1)]$_{\times 5}$ layers (see Figure S3 in the supplementary information), and will therefore affect the skyrmion density and size. The zero-field MFM images from Figs. \ref{fig:Fig4}(a)-(f) for these layers show regular maze domain patterns that become denser as the Fe thickness is increased. For the fields at which extended domains have been erased and solely skyrmions exist [Figs. \ref{fig:Fig4}(g)-(l)], this translates into an increasing skyrmion density and decreasing skyrmion size, in agreement with the observed trend for the MFM contrast shown with green symbols in Figure \ref{fig:Fig3} and also with previous observations \cite{Soumyanarayanan2017}. By looking at Figure \ref{fig:Fig4}(f), the domain pattern is not as regular and continous as for the other samples. In its corresponding skyrmion-only image from Figure \ref{fig:Fig4}(l), the skyrmions are not as easily discernable, but the ones that can be discerned have the smallest MFM contrast [$|\Delta f| = (0.25\pm0.1)\,$Hz from Figure \ref{fig:Fig3}] compared to the other samples. An interesting observation is that even though the FM layer with $x = 0.35\,$nm from Figure \ref{fig:Fig4}(l) does not show clear skyrmions, the corresponding trilayer sample from Figure \ref{fig:Fig2}(q) does show (mainly tubular) skyrmions very clearly. The skyrmions in the trilayer can only appear if skyrmions exist in the constituent FM layers. Note that the observed changes cover an Fe-sublayer thickness range that is varied in sub-\AA\, increments (see Methods also), making this structure extremely sensitive to changes in magnetic parameters while allowing a very fine tuning of the skyrmion density and size. 

To address the MFM observations in the single FM layers and ultimately connect with the trilayer results, we next discuss the exact implications of the Fe thickness variation on the magnetic parameters and DMI constant values in the [Ir(1)/Fe($x$)/Co(0.6)/Pt(1)]$_{\times 5}$ layers. More specifically, the DMI, exchange and anisotropy energies are contained within the material parameter $\kappa = \frac{\pi D}{4\sqrt{A K_\text{eff}}}$ \cite{Kiselev2011, Rohart2013, Leonov2016, Bogdanov1994, Heide2008}, where $D$ is the DMI constant, $A$ is the exchange stiffness and $K_\text{eff}$ is the magnetic perpendicular anisotropy (which is the same as the total effective magnetic anisotropy for the case of ultra-thin films \cite{Johnson1996, Lemesh2017, Wang2021}). The $K_\text{eff}$ values were extracted from in-plane M-H loops (see Figure S3 and corresponding text in the supplementary material) and are presented in Figure \ref{fig:Fig5}(a). We find that $K_\text{eff}$ decreases from 150\,$\pm$\,25\,kJ/m$^3$ to 7\,$\pm$\,3\,kJ/m$^3$ with increasing Fe thickness; this trend is also in agreement with other reports for similar systems \cite{Katayama1991, Chen2015, Soumyanarayanan2017}. Figure \ref{fig:Fig5}(b) shows the $D$ values which we find to vary between 1.7\,$\pm$\,0.2\,mJ/m$^{2}$ and 2.0\,$\pm$\,0.2\,mJ/m$^{2}$, with highest values for the thicker Fe layers. Note that both the anisotropy decreasing and the DMI increasing with increasing Fe thickness lower the domain wall energy and thus favor smaller domains in order to decrease the magnetostatic energy of the system. Being an interfacial effect, the DMI generally decreases with increasing FM layer thickness, just as the perpendicular magnetic anisotropy. Although the opposite behaviour is observed here, we attribute this to an incomplete Fe layer at the Ir interface even for our largest Fe-sublayer thickness. Therefore, it is expected that the DMI will increase as a more continuous Fe layer is formed. Note that our $D$ values are comparable to those reported in ref.\,\cite{Soumyanarayanan2017} for similar Fe and Co thicknesses. As a side note, in ref.\,\cite{Soumyanarayanan2017} [Ir(1)/Fe($x$)/Co(0.6)/Pt(1)] layers with 20 repeats were used instead of 5. Interestingly, for our case we observe an increase in $D$ and also an increase in $K_\text{eff}$ as the repetition number increases for a fixed Fe-sublayer thickness (see Figure S5 in the supplementary information for further details and also for MFM investigations on these samples). Having determined the experimental values for $K_\text{eff}$ and $D$, we can now calculate the parameter $\kappa$ of each FM layer. As previously established \cite{Rohart2013, Soumyanarayanan2017}, $\kappa$ is an indication of the skyrmion stability: for 0 $<$ $\kappa$ $<$ 1 metastable and isolated skyrmions can be stabilized, whereas for $\kappa$ $\geq$ 1 a stable and dense skyrmion lattice can exist in an applied field. Note that for the $\kappa$ calculations we have used an estimated value for $A$ of 15\,pJ/m \cite{Sampaio2013, Metaxas2007, Vidal-Silva2017, Wang2018} (see Section 3 in the supplementary information). The trend in $\kappa$ is plotted in Figure \ref{fig:Fig5}(a) and it is found to be increasing from $\approx$ 1 to $\approx$ 5, consistent with the evolution from less- to highly-dense skyrmion arrays as observed by MFM (Figure \ref{fig:Fig4}) with increasing Fe thicknesses. Since $D$ does not vary substantially between our samples, $K_\text{eff}$  appears to be the main reason for the increase in $\kappa$ and skyrmion density, as also observed for [Ir(1)/Fe($x$)/Co(0.6)/Pt(1)]$_{\times 20}$ \cite{Soumyanarayanan2017} and for Pt/Co/Ta(/MgO) multilayers \cite{Wang2021, Wang2019}.
\begin{figure}[t]
	\includegraphics[width=\textwidth]{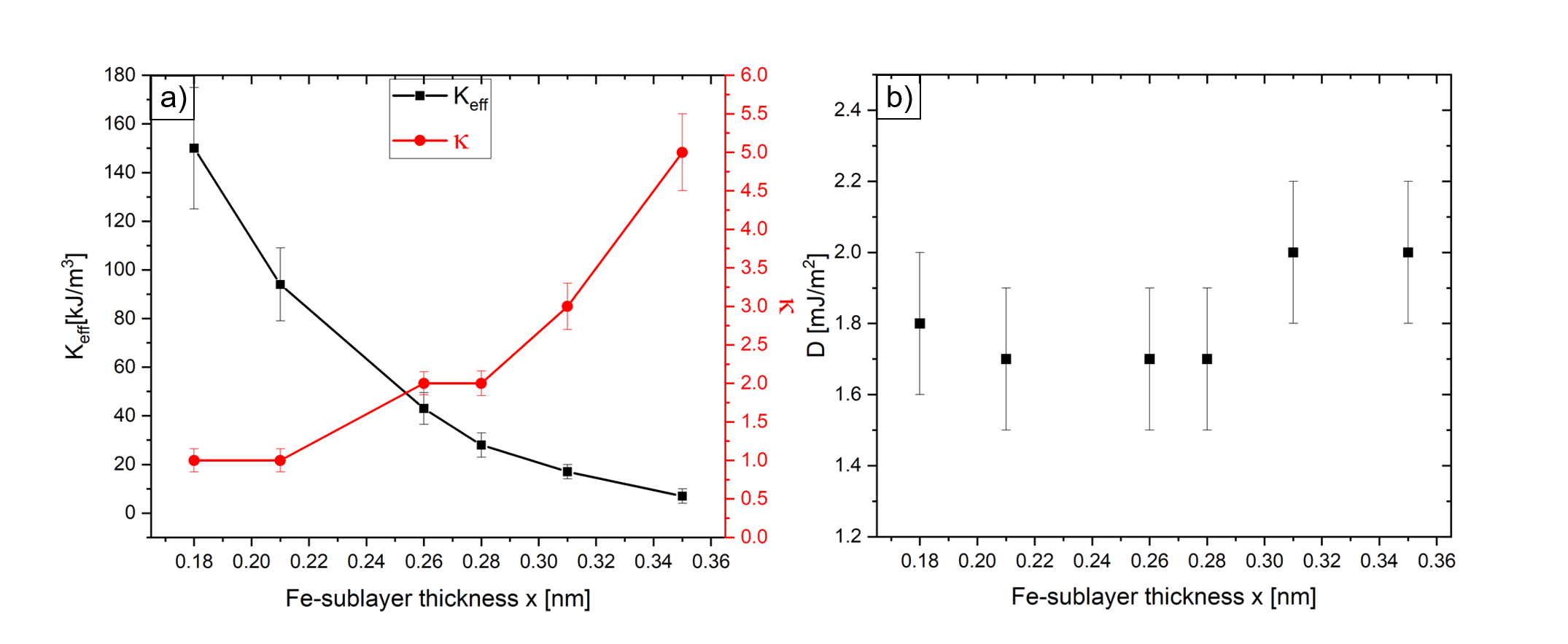}
	\caption{(a) Perpendicular magnetic anisotropy $K_\text{eff}$ and skyrmion stability parameter $\kappa$ values together with (b) DMI constants $D$ for the FM [Ir(1)/Fe($x$)/Co(0.6)/Pt(1)]$_{\times 5}$ layers as a function of the Fe-sublayer thickness.}
	\label{fig:Fig5} 
\end{figure}

Finally, since $K_\text{eff}$ and $D$ are known to play a key role in the skyrmion properties \cite{Soumyanarayanan2017, Wang2021, Wang2018, Wang2019}, the question is what combination of these parameters in the [Ir(1)/Fe($x$)/Co(0.6)/Pt(1)]$_{\times 5}$ layers gives rise to the observed coexistence or lack thereof for the incomplete and tubular skyrmions in the FM/FI/FM trilayer samples. To address this question, we performed micromagnetic simulations (see Methods) and the results are summarized in Table \ref{tab:Table1}. Note that: i) the FI alone supports very large (tens of microns) perpendicular domains and its DMI constant value has been set to $D_\text{FI}$ = +0.8\,mJ/m$^{2}$ \cite{Mandru2020}; ii) the DMI constant of the FM layers is actually negative, supporting clockwise N{\'e}el skyrmions \cite{Soumyanarayanan2017, Mandru2020}. By varying  $K_\text{eff}$, we determine for what $D$ values can the system support one, both or none of the two skyrmion types, identifying three main regimes: high, mid and low $K_\text{eff}$. In the high $K_\text{eff}$ regime, a uniform FM state, where no skyrmions are stable and the system behaves as a saturated ferromagnet, is obtained for $\lvert$$D$$\rvert$ values $\leq$\,2.2\,mJ/m$^{2}$. Since a high $K_\text{eff}$ is not favorable for skyrmion formation, having additionally $D$ values lower than a minimum value would also not permit any skyrmions to be stable, particularly with the FI layer supporting a larger domain size and thus a homogenuous magnetization. For larger $D$, i.e. 2.2\,$<$\,$\lvert$$D$$\rvert$\,$<$\,2.5\,mJ/m$^2$, an incomplete skyrmion becomes stable. Beyond an ever higher DMI threshold $\lvert$$D$$\rvert$\,$\geq$\,2.5\,mJ/m$^2$, the coexistence of incomplete and tubular skyrmions is achieved. In this case, due to the high $D$ of the FM layers and the existing $D_\text{FI}$, the skyrmions in the FM layers would also imprint a skyrmion in the FI layer, making tubular skyrmions energetically favorable along with the incomplete skyrmions. Finally, no $D$ value can stabilize the tubular skyrmion as a single phase. The existence of DMI threshold values is a clear signature that the skyrmion types are sustained by the DMI in the high $K_\text{eff}$ regime, i.e. DMI dominates over the dipolar interactions. In the mid $K_\text{eff}$ regime, the coexistence of tubular and incomplete skyrmions occurs beyond a DMI threshold $\lvert$$D$$\rvert$\,$\geq$\,1.4\,mJ/m$^2$, lower than the coexistence $D$ value for the high $K_\text{eff}$ regime. Since a lower $K_\text{eff}$ is favorable for skyrmion formation, $D$ can also be lower in this case, thus indicating the increasing effect of the dipolar interactions on skyrmion stabilization. In fact, as additional evidence, incomplete skyrmions are no longer stable, whereas we observe the existence of (magnetostatically stabilized) tubular skyrmions only for $\lvert$$D$$\rvert$ values lower than 1.4\,mJ/m$^2$. Eventually, in the low $K_\text{eff}$ regime, only the coexistence of tubular and incomplete skyrmions is observed independent of the DMI value, with dipolar interactions now dominating over the DMI. 

Since our experimental values are between 150\,$\pm$\,25\,kJ/m$^3$ and 7\,$\pm$\,3\,kJ/m$^3$ for $K_\text{eff}$ and between 1.7\,$\pm$\,0.2\,mJ/m$^{2}$ and 2.0\,$\pm$\,0.2\,mJ/m$^{2}$ for $D$, it is now clear why we find that all of our samples fall within the coexistence range, although approaching an incomplete state only and a tubular state only for small and large Fe thicknesses, respectively [see Figure \ref{fig:Fig3} and Figure \ref{fig:Fig2}]. Even though we already reach the right $K_\text{eff}$ values to allow the stabilization of tubular skyrmions only, none of the samples have $D$ lower than 1.5\,mJ/m$^{2}$ (taking into account the error in the BLS measurements). Regarding the stabilization of an incomplete skyrmion, even though in principle we could experimentally obtain an even higher $K_\text{eff}$ than 150\,$\pm$\,25\,kJ/m$^3$ by slightly decreasing the Fe-sublayer thickness to below 0.18\,nm, most likely the $D$ constant in this case will still be comparable to the 0.18\,nm sample, thus hindering the stabilization of incomplete skyrmions only.
\begin{table}[h]
	\centering
	\resizebox{\columnwidth}{!}{%
		\begin{tabular}{l|c|c|cl}
			& \multicolumn{1}{l|}{\textbf{\,\,\,\,\,\,\,high $K_\text{eff}$ (230\,kJ/m$^3$)}} & \multicolumn{1}{l|}{\textbf{\,mid $K_\text{eff}$ (55\,kJ/m$^3$)\,}} & \multicolumn{1}{l}{\textbf{\,low $K_\text{eff}$ (-10\,kJ/m$^3$)\,}} &  \\ \cline{1-4}
			\textit{uniform FM}      & $\lvert$$D$$\rvert$\,$\leq$\,2.2\,mJ/m$^2$                                    & $\boldmath\times$                                                       & $\boldmath\times$                                                        &  \\
			\textit{incomplete only\,} &\,2.2\,mJ/m$^2$\,$\textless$\,$\lvert$$D$$\rvert$\,$\textless$\,2.5\,mJ/m$^2$\,                                            & $\boldmath\times$                                                       & $\boldmath\times$                                                        &  \\
			\textit{coexistence}     &\,$\lvert$$D$$\rvert$\,$\geq$\,2.5\,mJ/m$^2$                                            &\,$\lvert$$D$$\rvert$\,$\geq$\,1.4\,mJ/m$^2$                           &\,$\lvert$$D$$\rvert$\,$\geq$\,0                                                  &  \\
			\textit{tubular only}    & $\boldmath\times$                                                          &\,$\lvert$$D$$\rvert$\,$\textless$\,1.4\,mJ/m$^2$                                & $\boldmath\times$  
			& 
		\end{tabular}
	}
	\caption{Summary of results obtained from micromagnetic simulations performed in 130 mT covering a large range of $K_\text{eff}$ values. Four different states (uniform FM, incomplete skyrmions only, coexisting skyrmions, and tubular skyrmions only) can be stabilized depending on the combination of $K_\text{eff}$ and $D$ values of the [Ir(1)/Fe($x$)/Co(0.6)/Pt(1)]$_{\times 5}$ layers. The $\boldmath\times$ symbol indicates the type of skyrmion that is not stable.}
	\label{tab:Table1}
\end{table} 
 
The stability of the different skyrmion types depends on the competition among magnetostatic, exchange, anisotropy and DMI energies, toghether with the interlayer exchange coupling of the FM layers (via the Pt and Ir interfaces) with the FI layer. The micromagnetic simulations also provide insight into the size and chirality of the different skyrmions types as determined by these competing energies. Cross-sections of the trilayer structure for each simulation result discussed above together with further details are given in Figure S6 of the supplementary information and corresponding text. As a final note, simulations were also performed for very high negative $K_\text{eff}$ values of -176 and -231\,kJ/m$^{3}$ (not shown), finding that for any $D$ value, the system does not support skyrmions, but stripe (for the case of tubular skyrmions) and vortex-like (for the incomplete skyrmions) states, which is compatible with what we would expect for in-plane systems.
\section{SUMMARY AND OUTLOOK}
In summary, we have developed trilayers consisting of skyrmion-hosting Ir/Fe/Co/Pt ferromagnetic multilayers separated by a ferrimagnetic layer. Such structures can simultaneously host two types of skyrmions: incomplete - existing in the Ir/Fe/Co/Pt layers only, and tubular - present throughout the whole film. By keeping the properties of the ferrimagnetic layer fixed, we have demonstrated that by changing the magnetic properties of the skyrmion-hosting layers, the coexistence of the two types of skyrmions can be steered towards a single type, of either incomplete or tubular skyrmions. The experimental findings were further explored by micromagnetic simulations, establishing the right combinations of perpendicular magnetic anisotropy $K_\text{eff}$  and DMI constants $D$ needed in the Ir/Fe/Co/Pt layers to stabilize the two skyrmion types either separately or collectively. The stabilization of purely incomplete skyrmions takes place for very high $K_\text{eff}$ and high $D$ values. Stabilizing a purely tubular skyrmion is possible for both lower $K_\text{eff}$ and $D$. The coexistence of incomplete and tubular skyrmions can occur for the whole range of $K_\text{eff}$ and for a wide range of $D$ values. Such heterostructure engineering as we show here, allows to effectively control the skyrmion type and density that can be obtained within the same material and could have useful implications in designing future skyrmion-based devices. Tuning the anisotropy may be achieved by irradiation {\cite{Fassbender2004}, hydrogen chemisorption/desorption \cite{Chen2022}, or magnetoionics \cite{Nichterwitz2021} which can be applied locally on the same structure. Such precise control permits the fabrication of, for example, racetrack structures where two input branches, one supporting tubular and the other incomplete skyrmions, are merged into a single output branch that equally supports both skyrmion types.

\section{Supporting Information}

Reproducibility control of FI layers by IP and OOP magnetormetry measurements (Figure S1); IP and OOP VSM loops for Trilayer heterostructures (Figure S2); IP and OOP magnetization curves for FM layers (Figure S3); Schematical representation of BLS measurement geometry (Figure S4); Magnetometry and MFM comparison of 5, 10 and 20 repeats FM layer with Fe=0.28 nm thickness (Figure S5); Spin cross-sections obtained from micromagnetic simulations (Figure S6)

\section{ACKNOWLEDGMENTS}
O.Y. and A.-O.M. thank Empa for financial support. P.M.V. and T.D. acknowledge the European Union’s Horizon 2020 research and innovation programme under the Marie Skłodowska-Curie grant agreement number 754364. This work has also been supported by the Project entitled “The Italian factory of micromagnetic modeling and spintronics” No. PRIN 2020LWPKH7 funded by the Italian Ministry of University and Research. 
\section{Methods}
\subsection{Magnetic force microscopy measurements}
The MFM measurements were performed using a home-built high-vacuum ($\approx$ 10$^{-6}$\,mbar) system equipped with an in-situ out-of-plane magnetic field of up to $\approx$ 300\,mT. By operating the MFM in vacuum, we obtain a mechanical quality factor $Q$ for the cantilever of $\approx$ 200k, which increases the sensitivity by a factor of $\approx$ 40 compared to MFM performed under ambient conditions, and also allows thin magnetic coatings on the tip to minimze the influence of its stray field on the micromagnetic state of the sample. SS-ISC cantilevers from Team Nanotech GmbH with a tip radius below 5\,nm (without any coating) were used. In order to make the cantilever tip sensitive to magnetic fields, we coated the tip with a Ta(2\,nm)/Co(3\,nm)/Ta(4\,nm) layer at room temperature using DC magnetron sputtering. A Zurich Instruments phase locked loop (PLL) system was used to oscillate the cantilever on resonance at a constant amplitude of $A_{\rm rms} = 5\,$nm and to measure the frequency shift arising from the tip-sample interaction force derivative. Note that the frequency shift is negative (positive) for an attractive (repulsive) force (derivative). For the MFM data shown in Figs. 2, 3 and in Figure S5, an up field was applied and an MFM tip with an up magnetization was used. Therefore, the skyrmions have a down magnetization just as the ones in the micromagnetic simulations (Figure S6). The up tip magnetization and the down magnetization of the skyrmions then generates a positive frequency shift contrast. Note that in order to quantitatively compare the MFM contrasts from different samples, we used the frequency-modulated distance feedback method described in ref.\,\cite{Zhao2018}. This method allows keeping the tip-sample distance constant with a precision of $\approx$ 0.5\,nm over many days, even after re-approaching the same tip on different samples and in applied magnetic fields. All this can be achieved without ever bringing the tip in contact with the sample surface such that the magnetic coating of the tip remains intact and therefore the same tip can be used for all samples.
\subsection{Sample preparation}
All films were grown using DC magnetron sputtering under a 2\,{\textmu}bar Ar atmosphere using an an AJA Orion-8 system with base pressure of ${\approx}$ 1\,$\times$\,10$^{-9}$\,mbar. All multilayers were deposited onto thermally oxidized Si(100) substrates coated with Ta(3\,nm)/Pt(10\,nm) seed layers; after growth, all films were capped with Pt(6\,nm) for oxidation protection. The substrates were annealed at $\approx$ 100\,$^{\circ}$C for an hour and cooled down until they reached temperatures close to room temperature prior to each deposition. The thickness for each sublayer was determined by regular calibrations performed using X-ray reflectivity on samples containing single layers of each individual element. Since the [Ir(1)/Fe($x$)/Co(0.6)/Pt(1)]$_{\times 5}$ FM layers are particularly sensitive to the Fe and Co thicknesses, the sub-\AA\, variation in the Fe thickness $x$  was controlled by alternating the deposition time between 29\,s and 56\,s from $x$ = 0.18\,nm to 0.35\,nm, respectively, while keeping the shutter opening/closing time $\textless$\,1\,s. Moreover, the reproducibility of FM layers was verified periodically by re-growing a FM sample with $x = 0.28\,$nm Fe thickness and performing MFM measurements under the same conditions as for the target samples. Similarly, the reproducibility of the FI layer was also tested by re-growing the same layer periodically and performing magnetometry measurements (see also Section 1 in the supplementary information).
\subsection{Magnetometry measurements}
The bulk magnetic properties of all samples were determined by vibrating sample magnetometry (VSM) using a 7\,T Quantum Design system. The measurements were performed at 300\,K for in-plane (IP) and out-of-plane (OOP) sample geometries and in fields of up to 2\,T. All samples were measured using the same VSM holders (i.e. one dedicated for IP and another one for OOP measurements). In addition, the background signal coming from the VSM holder and bare Si substrates was periodically checked \cite{Mandru2020-VSM} to ensure a clean magnetic signal coming from only the trilayers or from the individual FM and FI layers.
\subsection{Brillouin light scattering measurements}
The effective DMI constants of the FM layer samples were determined using BLS measurements from thermally excited spin waves ($k$) in the backscattering geometry. A monochromatic laser beam that is set to 150\,mW power (having wavelength $\lambda$ = 532\,nm) was focused on the sample surface through a camera objective of numerical aperture NA = 0.24. The scattered light was frequency analyzed by a Sandercock-type (3\,+\,3)-tandem Fabry-Perot interferometer \cite{Mock1987}. An in-plane field sufficient to saturate all samples ($\mu_\text{0}H$ = $\pm$0.7\,T) was applied along the z-axis, while the incident light ($k_{i}$) was swept along the perpendicular direction ($x$-axis), corresponding to the Damon-Eshbach (DE) geometry [see section 4 in the supplementary information and corresponding figures for further details]. 
\subsection{Micromagnetic simulations details}
The micromagnetic computations were carried out by means of a state-of-the-art micromagnetic solver PETASPIN \cite{Giordano2012} based on the finite difference scheme and which numerically integrates the Landau-Lifshitz-Gilbert (LLG) equation by applying the Adams-Bashforth time solver scheme:   
\begin{equation}
	\frac{d{\bf m}}{d\tau} = -({\bf m} \times {\bf h}_{\rm eff}) + \alpha _{\rm G} \left( {\bf m} \times \frac{d{\bf m}}{d\tau} \right)\,,
	\label{Giovannis eq}
\end{equation}
where $\alpha _{\rm G}$ is the Gilbert damping, ${\bf m } = {\bf M} / M_{\rm s}$ is the normalized magnetization, and $\tau = \gamma _0 M_{\rm s} t$ is the dimensionless time, with $\gamma _0$ being the gyromagnetic ratio, and $M_{\rm s}$ the saturation magnetization. $ {\bf h}_{\rm eff} $ is the normalized effective field in units of $M_{\rm s}$, which includes the exchange, interfacial DMI, magnetostatic, anisotropy and external fields \cite{Li2019,Tomasello2014}. The DMI is implemented as:
\begin{equation}
	\epsilon_{\rm InterDMI} = D\left[ m_z \nabla \cdot {\bf m} - ({\bf m} \cdot \nabla) m_z \right]\,.
	\label{Dieters eq}
\end{equation}

The [Ir(1)/Fe($x$)/Co(0.6)/Pt(1)]$_{\times 5}$ FM layers are simulated by 5 repetitions of a 1\,nm-thick CoFe ferromagnet separated by a 2\,nm-thick Ir/Pt non-magnetic layer. Each FM layer is coupled to the other ones by means of the magnetostatic field only (exchange decoupled); for simplicity, we neglect any Ruderman-Kittel-Kasuya-Yosida (RKKY) interactions. For the FM layers, we used the experimentally-obtained (except for the exchange constant $A$, which was kept at 15\,pJ/m for all simulations) physical parameters given in the inset tables from Figure S3. The ferrimagnetic [(TbGd)(0.2)/Co(0.4)]$_{\times 6}$/(TbGd)(0.2)] multilayer is simulated by a 4\,nm-thick magnetic layer, with physical parameters $M_\text{S} = 488$\,kA/m and $K_\text{u} = 486$\,kJ/m$^3$; an exchange constant $A$ = 4\,pJ/m was used in agreement with our prior work for rare-earth-transition metal alloy layers \cite{Zhao2019}. We use a discretization cell size of $3\,\times\,3\,\times\,1\,$nm$^3$. The top ferromagnetic layer of the bottom FM layer is coupled to the first 1\,nm of the ferrimagnetic layer via an RKKY-like interlayer exchange coupling \cite{Tomasello2017}. The bottom Ir/Fe/Co/Pt layer finishes with a 1\,nm-thick Pt layer that is known to lead to a large RKKY exchange coupling to the FI layer; we set a positive value of the constant (ferromagnetic coupling) equal to $0.8\,$mJ/m$^2$ from \cite{Omelchenko2018}. The top FM layer grown on the top of the FI layer starts with a 1\,nm-thick Ir layer. Since the RKKY coupling through 1\,nm of Ir is known to be very weak \cite{Meijer2020}, it has been neglected here. In all the simulations, an out-of-plane external field $H_{\rm ext} = 130$\,mT is applied.\\

\clearpage

\
\end{document}


\title{Supporting Information: Tuning the coexistence regime of incomplete and tubular skyrmions in ferro/ferri/ferromagnetic trilayers}

\author{O\u{g}uz Y{\i}ld{\i}r{\i}m}
\email{oguz.yildirim@empa.ch}
\affiliation{Empa, Swiss Federal Laboratories for Materials Science and Technology, CH-8600 Dübendorf, Switzerland}
\author{Riccardo Tomasello}
\affiliation{Department of Electrical and Information Engineering, Politecnico di Bari, 70125 Bari, Italy}
\author{Yaoxuan Feng}
\affiliation{Empa, Swiss Federal Laboratories for Materials Science and Technology, CH-8600 Dübendorf, Switzerland}
\author{Giovanni Carlotti}
\affiliation{Dipartimento di Fisica e Geologia, Università di Perugia, 06123 Perugia, Italy}
\author{Silvia Tacchi}
\affiliation{Istituto Officina dei Materiali - IOM, 06123 Perugia, Italy}
\author{Pegah Mirzadeh Vaghefi}
\affiliation{Empa, Swiss Federal Laboratories for Materials Science and Technology, CH-8600 Dübendorf, Switzerland}
\author{Anna Giordano}
\affiliation{Department of Mathematical and Computer Sciences, Physical Sciences and Earth Sciences, University of Messina, I-98166 Messina, Italy}
\author{Tanmay Dutta}
\affiliation{Empa, Swiss Federal Laboratories for Materials Science and Technology, CH-8600 Dübendorf, Switzerland}
\author{Giovanni Finocchio}
\affiliation{Department of Mathematical and Computer Sciences, Physical Sciences and Earth Sciences, University of Messina, I-98166 Messina, Italy}
\author{Hans J. Hug}
\affiliation{Empa, Swiss Federal Laboratories for Materials Science and Technology, CH-8600 Dübendorf, Switzerland}
\affiliation{Department of Physics, University of Basel, CH-4056 Basel, Switzerland}
\author{Andrada-Oana Mandru}
\email{andrada-oana.mandru@empa.ch}
\affiliation{Empa, Swiss Federal Laboratories for Materials Science and Technology, CH-8600 Dübendorf, Switzerland}

\clearpage
\pagenumbering{arabic} 
\newpage


\maketitle 
\clearpage
\section{Section 1: Reproducibility of the FI layers}
\begin{figure}[htbp]
	\includegraphics[width=\textwidth]{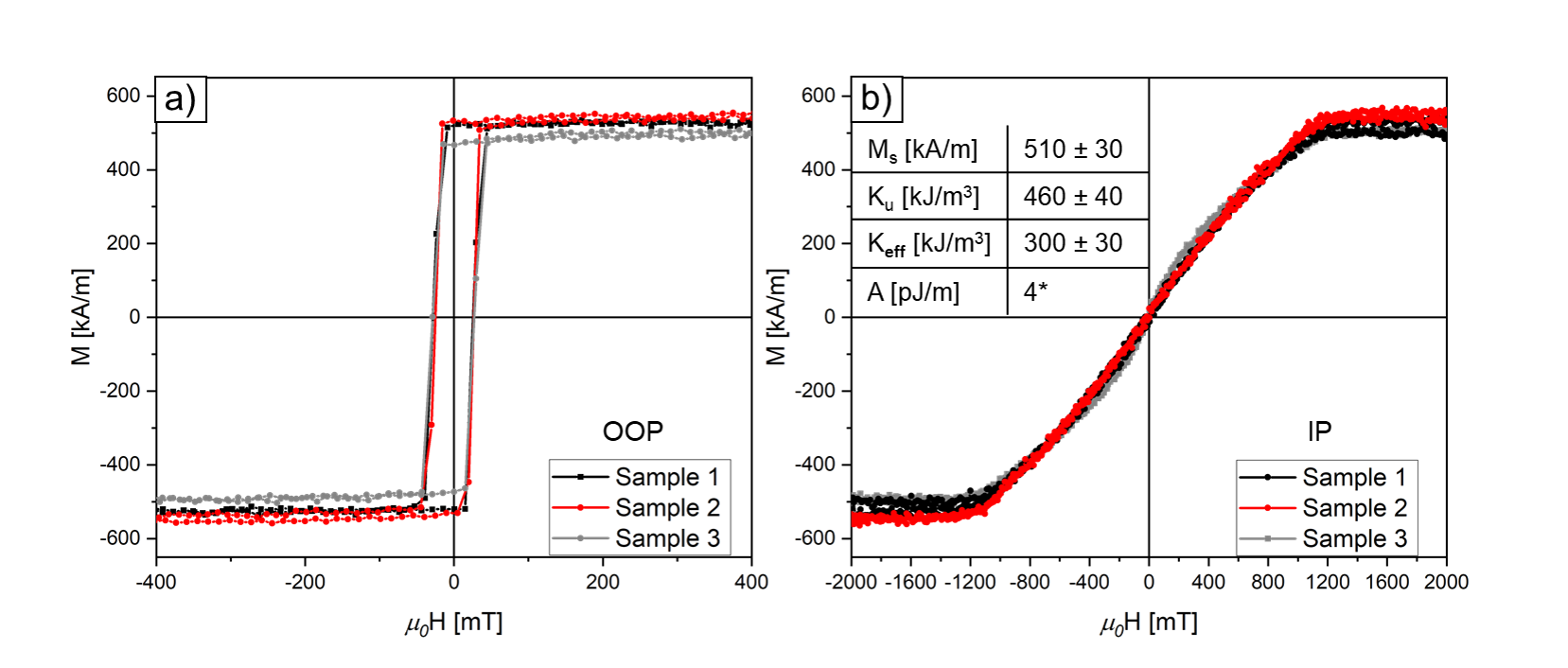}
	\caption{(a) out-of-plane (OOP) and (b) in-plane (IP) magnetization measurements of three FI [(TbGd)(0.2)/Co(0.4)]$_{\times 6}$/(TbGd)(0.2)] layers grown several samples apart. The inset in (b) shows the magnetic parameters calculated from the magnetization loops (* except for $A$, which is taken from ref.\,\cite{Zhao2019}).}
	\label{fig:FigS1} 
\end{figure}
The reproducibility of the ferrimagnetic (FI) layers was tested throughout this study by periodically re-growing them and performing magnetometry measurements. Given in Figs. \ref{fig:FigS1}(a) and (b) are the out-of-plane (OOP) and in-plane (IP) magnetization measurements performed on three different FI samples, demonstrating that the magnetic properties of the FI layers can be considered as constant for all trilayer samples prepared for this study. The minute differences that can be seen from the magnetization curves remain within the uncertainity levels given in the table shown as an inset in Figure \ref{fig:FigS1}(b). We also note that the magnetic properties of the FI layers in this study are very close to the values of the FI in our previous investigations \cite{Mandru2020}, i.e. $M_\text{S}$ = 490\,$\pm$\,30\,kA/m $K_\text{eff}$ = 340\,$\pm$\,30\,kJ/m$^3$.
\clearpage

\section{Section 2: Magnetic properties of the trilayer heterostructures}
\begin{figure}[htbp]
	\includegraphics[width=\textwidth]{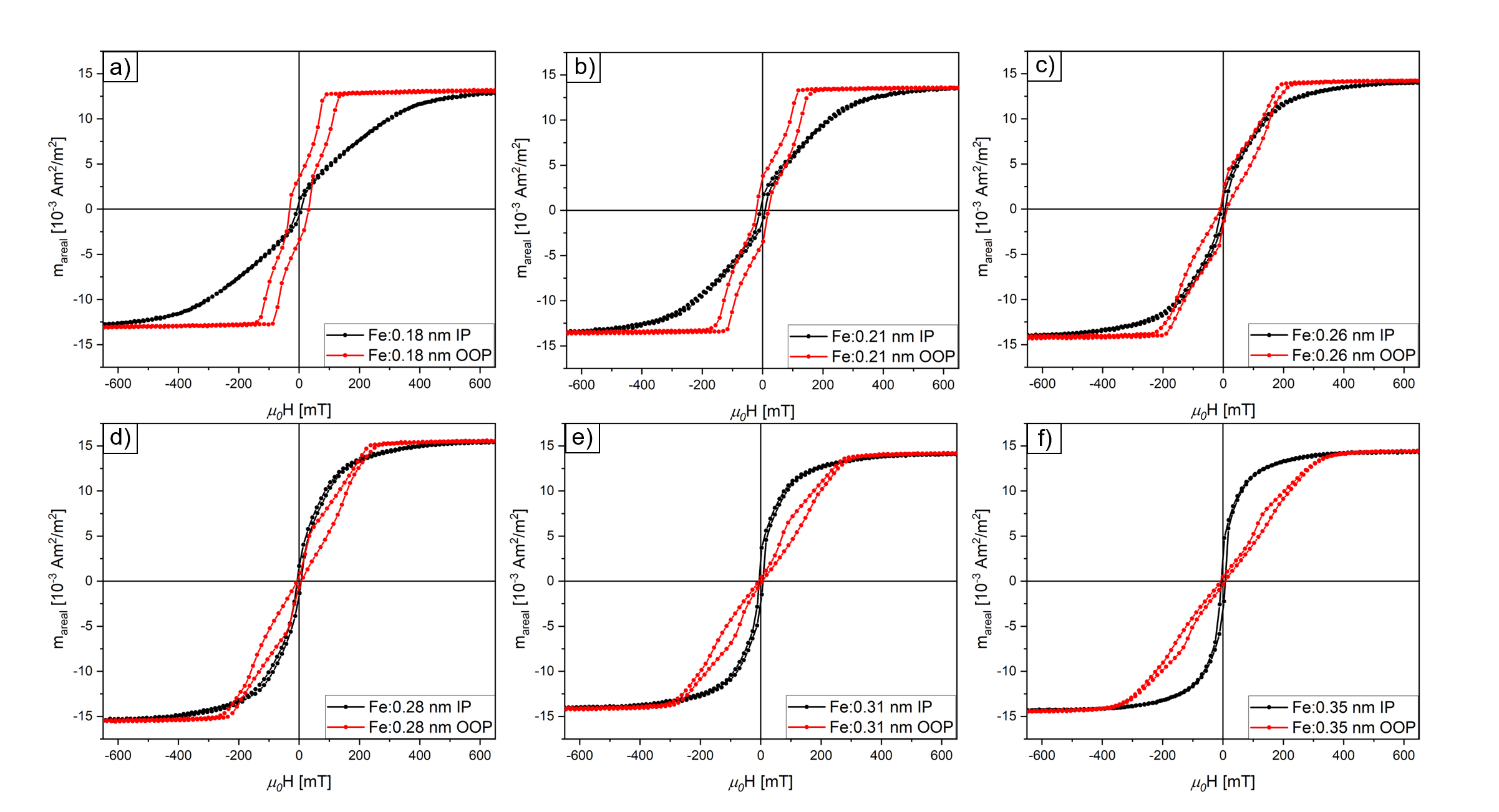}
	\caption{(a)-(f) IP (black) and OOP (red) hysteresis measurements of the trilayer heterostructures consisting of a FI layer sandwiched between two [Ir(1)/Fe($x$)/Co(0.6)/Pt(1)]$_{\times 5}$ FM layers, with the Fe-sublayer thickness $x$ varied from 0.18\,nm to 0.35\,nm. Since these loops represent the sum of all three layers (each with different magnetic moment), the $y$-axis is given in units of magnetic moment per unit are and not in units of magnetization. Corresponding FM-only and FI-only layer measurements are given in Figs. \ref{fig:FigS3} and  \ref{fig:FigS1}, respectively.}
	\label{fig:FigS2} 
\end{figure}
\clearpage

\section{Section 3: Magnetic properties of the FM layers}
\begin{figure}[htbp]
	\includegraphics[width=\textwidth]{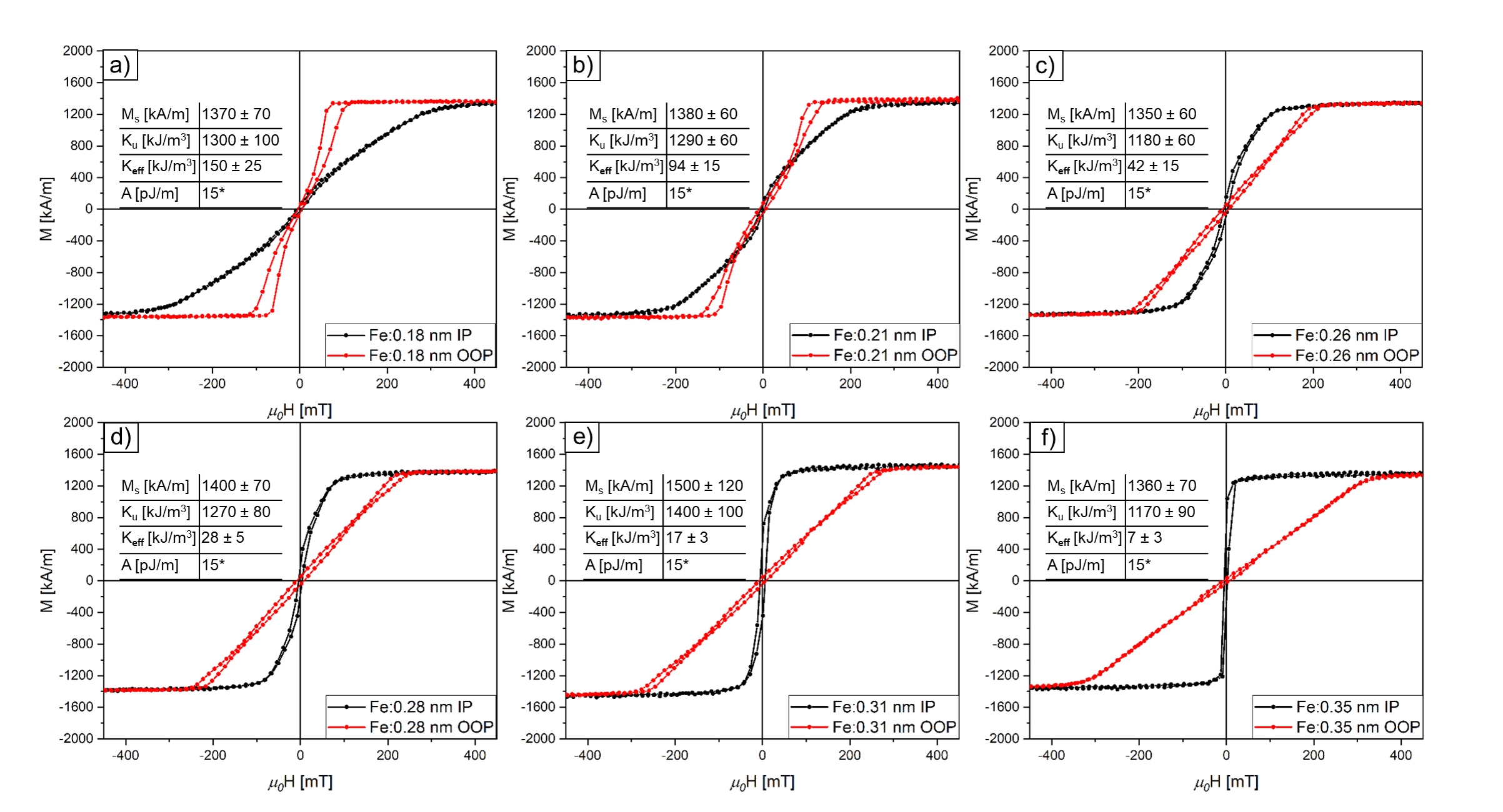}
	\caption{(a)-(f) IP (black) and OOP (red) magnetization curves of the individual [Ir(1)/Fe($x$)/Co(0.6)/Pt(1)]$_{\times 5}$ FM samples with varied Fe thickness $x$. Insets in each graph show the magnetic parameters extracted from the corresponding measurement (* except for $A$, which is taken from refs. \cite{Sampaio2013, Metaxas2007, Vidal-Silva2017, Wang2018}).}
	\label{fig:FigS3} 
\end{figure}
Figure \ref{fig:FigS3} displays the IP (black) and OOP (red) magnetometry measurements of the FM [Ir(1)/Fe($x$)/Co(0.6)/Pt(1)]$_{\times 5}$ layers for all six investigated Fe thicknesses. Inside each graph as insets are the magnetic parameters of each sample extracted from the corresponding data. Increasing the Fe thickness within the investigated range does not have a major impact on the saturation magnetization $M_\text{S}$, as the variations mostly stay within the uncertainity. The anisotropy field $H_{A}$ (saturation field for the in-plane loop) on the other hand decreases with increasing Fe thickness. By using the $M_\text{S}$ and $H_{A}$ values obtained from the magnetometry data, $K_\text{eff}$ values were calculated using the following equation \cite{Johnson1996, Lemesh2017, Wang2021}:
\begin{equation}
	K_\text{eff} = \frac{1}{2}\mu_\text{0} H_{A} M_\text{S}\,,
		\label{Eq:Eq1}
\end{equation}
where $\mu_\text{0}$ is the permeability of vacuum. Note that the magnetic perpendicular anisotropy that should technically appear in the $\kappa$ formula in the main text is $K = K_\text{u} - \frac{1}{2}\mu_\text{0} M_\text{S}^{2}$, with $K_\text{u}$ being the uniaxial magnetic anisotropy and $\frac{1}{2}\mu_\text{0} M_\text{S}^{2}$ the magnetostatic energy. However, the magnetic perpendicular anisotropy $K$ is the same as the total effective magnetic anisotropy $K_\text{eff}$ given in eq.\,(1) above for the case of ultra-thin films \cite{Johnson1996, Lemesh2017} and that is why we use the notation $K_\text{eff}$ in the main text.

Regarding the exchange stiffness constant $A$, typically, for similar systems, it is estimated to be around 15\,pJ/m \cite{Sampaio2013, Metaxas2007, Vidal-Silva2017, Wang2018} and this is also the value that we used in this study for both the $\kappa$ calculations and the micromagnetic simulations. However, since all of the magnetic parameters used in this study were experimentally obtained, we have also attempted to estimate $A$ for each FM layer by fitting $M_\text{S}$\,vs.\,T  measurements (with T varied from 10\,K to 400\,K) to Bloch's $T^{\frac{3}{2}}$ law \cite{Zeissler2017}. From these fits, we obtain $A$ values varying from 15.8\,pJ/m to 16.4\,pJ/m. Even though this rough aproximation may not be suitable for thin film systems \cite{Yastremsky2019}, whatever errors come from this method would be the same for every sample. Since the estimated $A$ values remain virtually unchanged with varying Fe-sublayer thickness, using the same value for all films is justified. In order to be consistent with our previous work \cite{Mandru2020} and with other studies, we chose to use the 15\,pJ/m value.
\clearpage

\section{Section 4: Determination of DMI constants from BLS measurements}
\begin{figure}[htbp]
	\includegraphics[width=\textwidth]{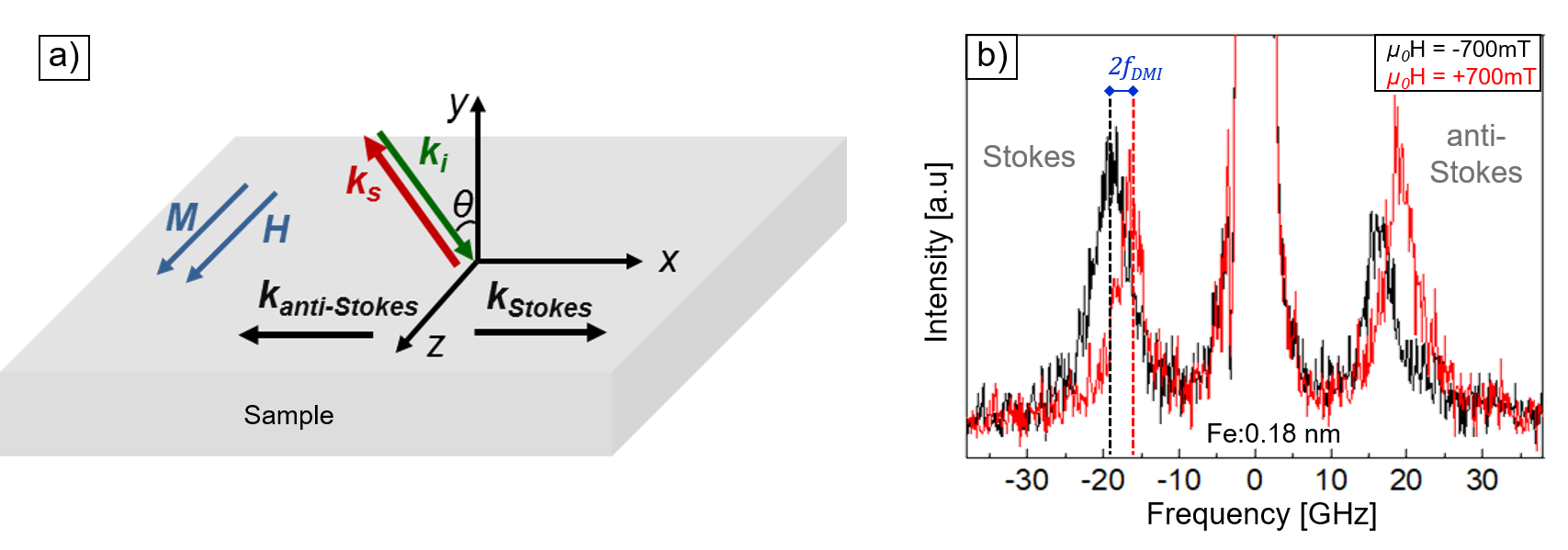}
	\caption{(a) Schematic diagram of the BLS measurement geometry, indicating the incident ($k_i$) and backscattered ($k_s$) light wave vectors, along with the magnetization and applied field directions; the spin waves propagating along the $x$-axis involved in the Stokes (anti-Stokes) process, i.e. in the generation (annihilation) of a magnon, are also shown. (b) Example BLS frequency spectra for the [Ir(1)/Fe($x$)/Co(0.6)/Pt(1)]$_{\times 5}$ FM sample with $x$ = 0.18\,nm obtained for two opposite field directions. Dashed lines highlight the peak positions for different field directions and are given as a guide to the eyes.}
	\label{fig:FigS4}
\end{figure}
In order to quantitatively estimate the magnetic parameters of the samples, we used the analytical expression of the spin waves in the DE configuration, valid for an in-plane magnetized FM film of thickness $t$ :
	
\begin{multline}
 f(k) = f_{0}(k)\,\pm\,f_{DMI}(k) = \\ 
 \frac{\gamma \mu_{0}}{2\pi}\sqrt{\left (H+Ak^{2}+\frac{M_{S}kt}{2}\right ).\left[H-\frac{2K_\text{u}}{M_{S}}+Ak^{2}+M_{S}\left(1-\frac{kt}{2}\right)\right]}\,\pm\,\frac{\gamma Dk}{\pi M_{S}}.
		\label{Eq:Eq2}
\end{multline}
where $K_\text{u}$ and $\gamma$ are the uniaxial magnetic anisotropy constant and the gyromagnetic ratio, respectively. These analyses were carried out fixing the exchange constant $A$ to the value of 15\,pJ/m and $M_{S}$ to values measured by VSM (Figure \ref{fig:FigS3}). The value of $\gamma$ is 176\,GHz/T, as obtained from the dependence of the average frequency $f_{0}(k)$ measured for different values of the applied magnetic field $H$. 

Then, for each sample we obtained the value of $D$ from the frequency asymmetry $f_{DMI}(k)$ between the Stokes and anti-Stokes peaks measured for $\mu_\text{0}H$ = $\pm$0.7\,T using $D = \frac{\pi M_{S}f_{DMI}}{\gamma k}$. The absolute values of $D$ for the different samples are reported in Figure 3(b) of the main text and the error reflects the uncertainties in $M_{S}$ and in determining $f_{DMI}$. The measured sign of $f_{DMI}$, and therefore that of $D$, accounts for a favored clockwise (CW) domain wall chirality in our samples, that is in fact in agreement with previous results on Co films with a Pt overlayer. Note that the sign of $D$ depends on the convention chosen in relation with the definition of the DMI Hamiltonian and of the BLS interaction geometry. Here, consistent with previous studies of domain walls, a negative/positive $D$ corresponds to N{\'e}el domain walls having CW/couter-clockwise (CCW) chirality. However, one should take into account that, in most of the previous BLS studies, the opposite convention is adopted, as summarized in a recent review paper \cite{Kuepferling2022}.
\clearpage

\section{Section 5: Comparison between 5, 10 and 20 repeats FM layers}
\begin{figure}[htbp]
	\includegraphics[width=\textwidth]{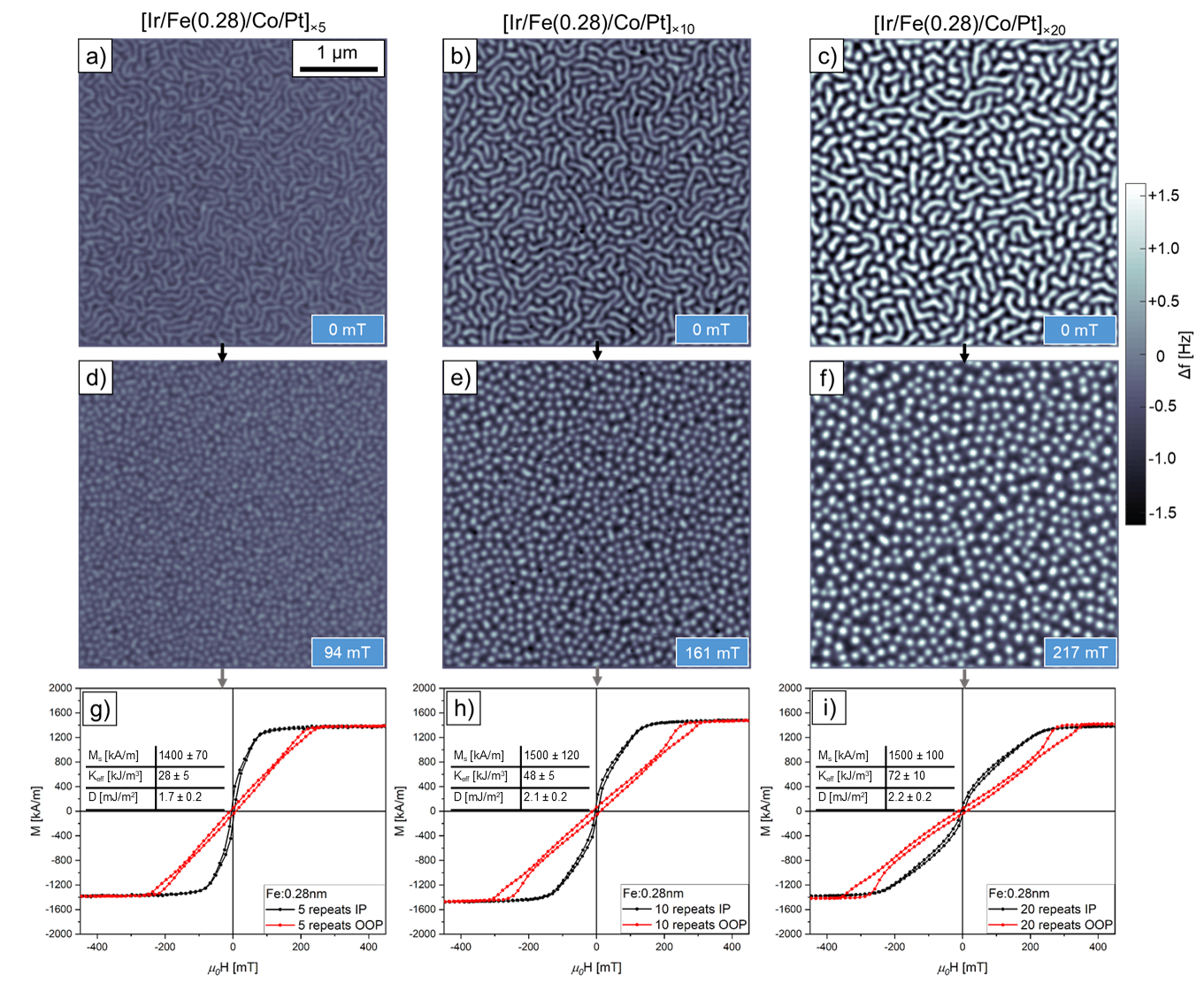}
	\caption{(a)-(c) 4\,{\textmu}m\,$\times$\,4\,{\textmu}m zero-field MFM images measured on FM [Ir(1)/Fe($x$)/Co(0.6)/Pt(1)]$_{\times N}$ layers with $N$ = 5, 10, and 20 for the same Fe thicknesses $x$ = 0.28\,nm; all images are taken at remanence after out-of-plane saturation in 450\,mT magnetic field. (d)-(f) Subsequent MFM images recorded at the fields where mainly skyrmions exist and no maze domains are left. The $|\Delta f|$ contrast scale is the same for all images. All images are taken with the same tip, at the same tip-sample distance and under the same conditions as the images from Figs. 2 and 3 in the main text. (g)-(i) Corresponding IP and OOP magnetometry loops with insets showing the relevant magnetic parameters for these samples.}
	\label{fig:FigS5} 
\end{figure}
\clearpage
We have also investigated the impact of the repetition number $N$ on [Ir(1)/Fe($x$)/Co(0.6)/Pt(1)]$_{\times N}$ FM layers for a fixed Fe-sublayer thickness $x$ = 0.28\,nm. A summary of the results is shown in Figure \ref{fig:FigS5} for $N$ = 5, 10, and 20. MFM investigations on these samples reveal that with increasing $N$, the skyrmion density decreases, consistent with an increase in $K_\text{eff}$ [see insets from Figs. \ref{fig:FigS5}(g)-(i)], just as for the [Ir(1)/Fe($x$)/Co(0.6)/Pt(1)]$_{\times 5}$ with varying Fe thickness described in the main text. The field that is required to attain a skyrmion-only state increases with increasing $N$, due mainly to the gain in magnetostatic energy, also consistent with the increased MFM $|\Delta f|$ contrast from 5 to 10 to 20 repeats [compare Figs. \ref{fig:FigS5}(d)-(f), all shown on the same contrast scale]. Interestingly, the $D$ value also increases with increasing $N$, from 1.7\,$\pm$\,0.2\,mJ/m$^{2}$ for $N$ = 5 to 2.2\,$\pm$\,0.2\,mJ/m$^{2}$ for $N$ = 20. These observations are in contrast to the Pt/Co/Ta/MgO system, where $D$ was found to be unaffected by the repetition number \cite{Wang2021} (note that here the $D$ constant was also determined by BLS). When comparing our values for $N$ = 20 with the identical [Ir(1)/Fe($x$)/Co(0.6)/Pt(1)]$_{\times 20}$ system from ref.\,\cite{Soumyanarayanan2017} for $x$ = 0.3\,nm, we find that our value of 2.2\,$\pm$\,0.2\,mJ/m$^{2}$ is larger than the one reported there, which is 1.9\,$\pm$\,0.2\,mJ/m$^{2}$.  However, $D$ was estimated in ref.\,\cite{Soumyanarayanan2017} by comparing domain periodicities from MFM with those from micromagnetic simulations; different methods of obtaining $D$ could lead to slightly different results. 
\clearpage

\section{Section 6: Further micromagnetic simulations results}
\begin{figure}[htbp]
	\includegraphics[width=0.97\textwidth]{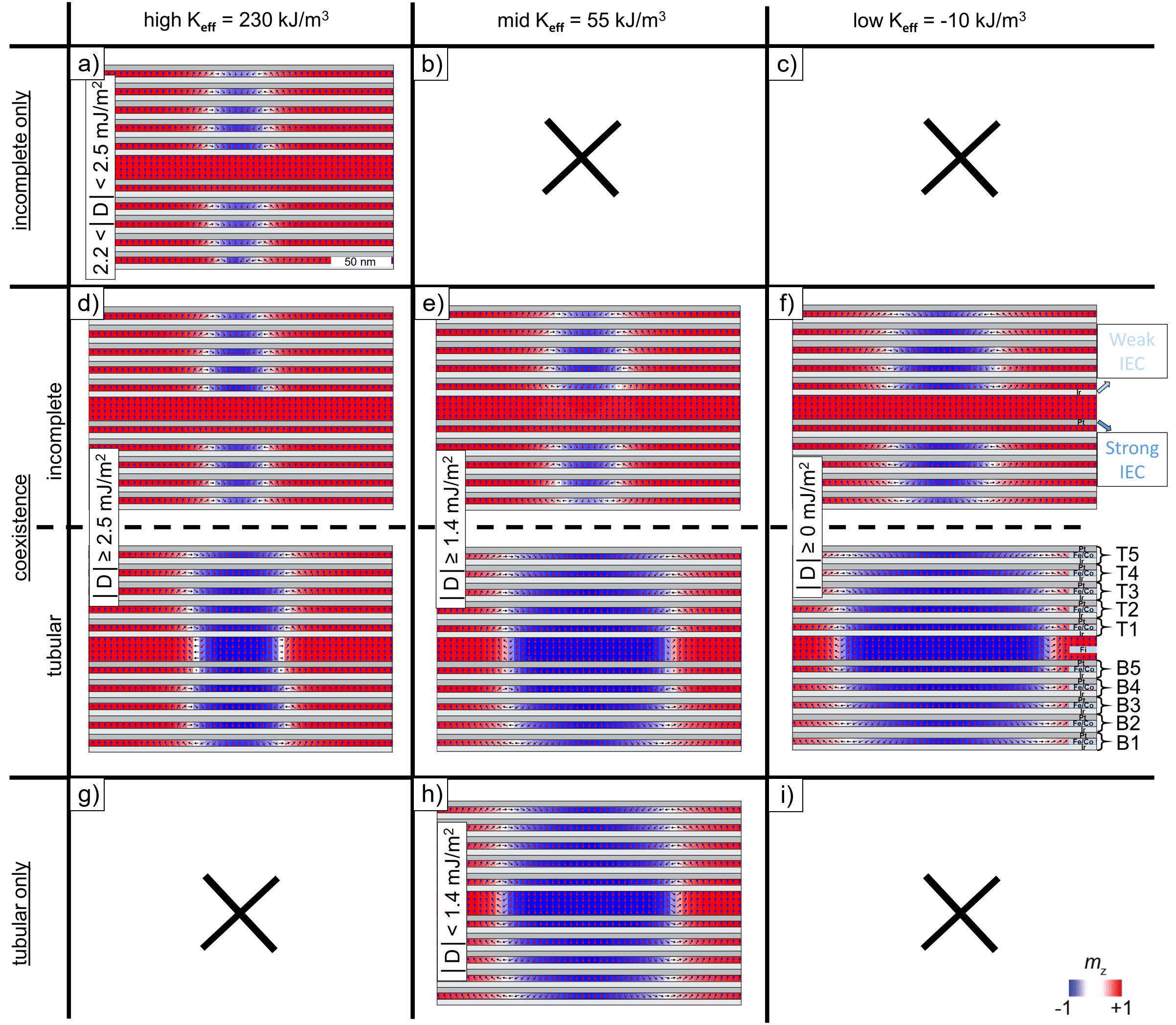}
\caption{Cross-sections of the trilayer heterostructures obtained from micromagnetic simulations for different $K_\text{eff}$ and $D$ values for the FM layers; the parameters for the FI layer remain fixed. The three main regions are divided according to the $K_\text{eff}$ used in the simulations: high $K_\text{eff}$ region (a),(d) and (g); mid $K_\text{eff}$ region (b), (e) and (h); and low $K_\text{eff}$ region (c), (f) and (i). The dashed line that divides the coexistence row separates the incomplete case (higher panel) from the tubular case (lower panel) and is meant as a guide to the eyes. The $\boldmath\times$ symbol indicates the type of skyrmion that is not stable. Individual sublayers T1-T5 and B1-B5 (lower panel) as well as the strength of the IEC through Ir and Pt (upper panel) are indicated in (f).}
\label{fig:FigS6}
\end{figure}
\clearpage
We perform systematic micromagnetic simulations for a range of uniaxial anisotropy constants $K_\text{u}$ and different (negative) interfacial DMI parameters $D$ of the FM layers. The magnetic parameters of the FI layer remain fixed and $D_\text{FI}$ is set to +0.8\,mJ/m$^{2}$. For each pair of $K_\text{u}$-$D$ values, we relax the system by starting the simulations from two different initial states, i.e. the incomplete skyrmion [clockwise (CW) N{\'e}el skyrmions in the FM layers, but no skyrmion is imposed in the FI] and the tubular skyrmion (again, CW skyrmions in the FM layers, but this time a skyrmion is also placed in the FI layer). In this way, we can confirm which state is stable. Even though $K_\text{u}$ is used in our modeling work (along with $M_{S}$, $D$ and $A$), we refer to the value of the total effective anisotropy constant $K_\text{eff}$ (as explained above in section 3 of the supplementary information) to facilitate the comparison with the experimental data. As already discussed in the main manuscript, we identify three regions, high, mid, and low $K_\text{eff}$ with corresponding $D$ values, leading to the stabilization of different skyrmion types. Shown in Figure \ref{fig:FigS6} are the cross-sections of the spin structures that can be stabilized in our system and already highlighted in Table I of the main text. 

In terms of size, we find that as $K_\text{eff}$ reduces, the diameter of both types of skyrmions increases. Note that from the MFM measurements and contrast analysis shown in the main text (Figs. 2 and 3), with decreasing $K_\text{eff}$ (or increasing Fe thickness) the skyrmion diameter becomes smaller, seemingly in contrast with the aforementioned simulation results. However, the simulations were all performed for a field of 130\,mT, whereas the fields required in the experiment to collapse all stripe domains and obtain a skyrmion-only magnetic state increase with decreasing $K_\text{eff}$, exceeding 130\,mT for some of the samples. For simulations performed in larger fields the skyrmion diameter will become smaller, in agreement with our experimental results.

We also explored the skyrmion chirality for the three $K_\text{eff}$ regimes and corresponding $D$ values of the FM layers. We first describe the case of high $K_\text{eff}$ and high $D$ shown in Figure \ref{fig:FigS6}(d), which is the same as our previous study \cite{Mandru2020}. For the case of incomplete skyrmions, the strong negative $D$ (in this case $\lvert$$D$$\rvert$\,$\geq$\,2.5\,mJ/m$^2$) favors CW skyrmions in all the Fe/Co sublayers of the top (T) and bottom (B) FM layers, except for the top-most sublayer of the bottom FM layer, i.e. B5; the skyrmion in this sublayer is supressed due to the large interlayer exchange coupling (IEC) between the FI (which in this case has a uniform magnetization) and the last Pt sublayer of the bottom FM layer. Note that a CW chirality is observed in all these sublayers due to the dominating effect of DMI over magnetostatics. For the tubular case, skyrmions now exist not only in sublayers B1-B4, but also in B5, albeit with a Bloch-type chirality arising from the competition between positive DMI in the FI, IEC through Pt and negative DMI in the FM layer. Sublayers B2-B4 have a CW chirality, whereas sublayer B1 now has a counter-clockwise (CCW) chirality, opposite to what is expected for negative $D$; we attribute this to an improved flux closure at the bottom layer that reduces the stray field below layer B1 and thus the overall magnetostatic energy. The chirality of the skyrmion that is now present in the FI is hybrid between CCW N{\'e}el and Bloch. Finaly, the chirality of the top Fe/Co sublayers T1-T5 remains CW (the IEC of the FI through the first Ir sublayer of the top FM layer is weaker than the IEC through Pt of the bottom FM layer). For the case of high $K_\text{eff}$ and lower $D$ shown in Figure \ref{fig:FigS6}(a), only the incomplete skyrmion can be stabilized and its chirality is the same as for the incomplete case in Figure \ref{fig:FigS6}(d). Considering the case of mid $K_\text{eff}$ and $D$ values $\lvert$$D$$\rvert$\,$\geq$\,1.4\,mJ/m$^2$, lower than the coexistence $D$ value for the high $K_\text{eff}$ regime, the contributions from magnetostatics becomes more pronounced. This leads to more complex spin textures of the skyrmions in the different sublayers of the top and bottom FM layers [compare Figs. \ref{fig:FigS6}(e) and \ref{fig:FigS6}(d)]. The incomplete skyrmion is now composed of two skyrmions with hybrid chiralities (some layers are CW and others CCW) and the tubular skyrmion has more bottom Fe/Co sublayers with a CCW chirality. For mid $K_\text{eff}$ but for $\lvert$$D$$\rvert$ values $\textless$\,1.4\,mJ/m$^2$ [Figure \ref{fig:FigS6}(h)], only a tubular skyrmion can be stabilized with a uniform CCW chirality in the bottom FM layer (dictated by the strong IEC through Pt) and a uniform CW chirality in the top FM layer (due to the magnetostatic field of bottom FM layer). This indicates that the IEC of the FI through Pt and also magnetostatics become more pronounced than the DMI contributions. For the last case, i.e. low $K_\text{eff}$ and any $D$ values shown in Figure \ref{fig:FigS6}(f), the incomplete skyrmion is again hybrid, but with more top and bottom CCW sublayers compared to the one in Figure \ref{fig:FigS6}(e). The tubular skyrmion on the other hand shows a very similar spin profile to that shown in Figure \ref{fig:FigS6}(h).

\clearpage